\documentclass[aps,graphicx,twocolumn,epsfig]{revtex4}
\usepackage{amsmath}
\usepackage{graphicx}

\begin{document}

\title{Beyond the Landau Criterion for Superfluidity}

\author{Sara Ianeselli$^1$, Chiara Menotti$^1$ and Augusto Smerzi$^{1,2}$}
\affiliation{
1) Istituto Nazionale di Fisica per la Materia BEC-CRS and
Dipartimento di Fisica, \\
Universit\`a di Trento, I-38050 Povo, Italy \\
2)Theoretical Division, Los Alamos National Laboratory, Los Alamos, 
NM 87545, USA  }

\date{\today}

\begin{abstract}

According to the Landau criterion for superfluidity, a Bose-Einstein
condensate flowing with a group velocity smaller than the sound
velocity is energetically stable to the presence of perturbing
potentials.  We found that this is strictly correct only for
vanishingly small perturbations. The superfluid critical velocity
strongly depends on the strength and shape of the defect. We
quantitatively study, both numerically and with an approximate
analytical model, the dynamical response of a one-dimensional
condensate flowing against an istantaneously raised spatially periodic
defect. We found that the critical velocity $v_c$ decreases by
incresing the strength of the defect $V_0$, up to to a critical value
of the defect intensity where the critical velocity vanishes.
\end{abstract}


\maketitle


{\it Introduction.} Dissipationless flow in homogeneous bosonic
systems is one of the manifestations of the interplay between
interatomic interaction and Bose-Einstein condensation. Bogoliubov
ground-breaking calculation of the many-body properties of a dilute
condensate showed that this interplay produces a linear dispersion of
the low-energy excitation modes which, according to Landau, inhibits
dissipation when the group velocity of the fluid is smaller than the
sound velocity \cite{Huang,pita,leggett}. 
 However, the question if a Bose-Einstein
condensed dilute gas (BEC) really displays this basic characteristics
of superfluidity remained open until its experimental realization in
1995.  Eventually, several typical superfluidity phenomena like
vortices with quantized circulation ~\cite{Abo-Shaeer, Hodby}, the
scissor mode ~\cite{Davis, Marago_scissors} and Josephson dynamics
\cite{jose} have been experimentally observed.  Instead, the direct
experimental verification of the Landau critical velocity has been
more elusive. The MIT group studied the response of a trapped
elongated condensate perturbed by a blue detuned laser beam
\cite{onofrio}.  They found a transition between the superfluid and
the dissipative regime at a laser velocity smaller than the estimated sound
velocity.  The reason for this discrepancy is not fully understood,
but it might be related to the topological creation of vortices
\cite{Jackson}.  It is worth noticing that even in superfluid $^4$He
the observed superfluid critical velocity is smaller than Landau's
prediction \cite{tilley}.

In this paper we study the response of a one dimensional homogeneous
condensate flowing with momentum $p$ against an instantaneously raised
defect.  The analysis is done numerically, solving the
Gross-Pitaevskii equation (GPE), and analytically, with both a
dynamical Bogoliubov approach and a nonlinear two-mode approximation.
The Bogoliubov linear framework provides a clear dynamical analysis of
the Landau criterion, predicting the transition from a superfluid to a
dissipative regime at a condensate group velocity equal to the sound
velocity.  However, and this is the main result of this paper, we
show that nonlinear corrections to the Landau criterion become very
important even for small strengths of the defect.

We consider, as a defect, a spatially periodic potential. This choice
is convenient for two reasons: first, in the linear regime it is
possible to selectively excite a given pair of quasi-particle modes,
and study their behaviour below and above the critical velocity;
second, periodic potentials can be easily taylored and accurately
controlled experimentally.

Our analysis has several analogies with the Bragg spectroscopy of 
condensates, which has been extensively investigated both
theoretically and experimentally
\cite{kozuma,Stenger,davidson,dalfovo,morgan}. 
The crucial difference is that we focus on the signature of the 
energetic instability associated with the Landau criterion of 
superfluidity, and on the study of how the nonlinear dynamics
modifies the Landau critical velocity.

In particular, for a spatially periodic defect, we find that the
Landau critical velocity $v_c$ rapidly decreases with increasing
defect strength, up to a critical value where superfluidity
disappears.  This result might be related to the downwards shift of
the excitations frequencies with respect to the Bogoliubov spectrum,
as predicted in \cite{dalfovo}.  While the regimes investigated in
Bragg spectroscopy experiments are still described with a good
accuracy by linear Bogoliubov theory, the experimental investigation
of our predictions would require Bragg experiments in highly
non-linear regime, i.e. beyond the regimes investigated so far. 
On the other hand, the investigation of energetical instability
in the linear regime of very weak defects, might require times
unaccessible experimentally.
 

{\it Bogoliubov Analysis.} We describe the BEC dynamics by an
effective one-dimensional GPE

\begin{eqnarray} \label{GPE_defect}
  i\hbar\dot{\psi}(x, t) 
=\left[-\frac{\hbar^2 \partial_x^2}{2m}
 +gn L |\psi|^2+ \Theta(t)
V_{def}(x) \right]\psi(x,t)
\end{eqnarray}
with $g=4\pi \hbar^2a/m$, $m$ the atomic mass, $a$ the $s$-wave
interatomic scattering length, $n$ the 3D density and $L$ the length
of the system. Moreover $V_{def}(x)$ is the defect istantaneously
raised at $t=0$ and $\Theta(t)$ the Heaviside theta function.  The GPE
can be linearized by writing the condensate wave-function, normalised
to $1$ in the interval $L$, as

\begin{eqnarray} \label{Bog_ansatz}
&& \psi(x, t) =\frac{1}{\sqrt{L}} e^{i(px-\mu t)/\hbar} \times \\
&\times& \left[1+ \sum_q \left( c_q(t) U_q e^{i (q x/\hbar-\omega t)} +
c^*_q(t) V_q^* e^{i (q x/\hbar + \omega t)} \right) \right] ,
\nonumber
\end{eqnarray}
being $\mu=p^2/2m+gn$ the chemical potential. The coefficients
$c_q(t)$ are time-dependent quasi-particle amplitudes and $U_q$, $V_q$,
normalized to $|U_q|^2-|V_q|^2=1$, satisfy the Bogoliubov equations
with dispersion

\begin{eqnarray}
\hbar\omega_q= \frac{pq}{m}+
\sqrt{\frac{q^2}{2m}\left(\frac{q^2}{2m}+2gn\right)}.
\end{eqnarray}
The momentum $q$ represents the momentum of the quasi-particle
excitation with respect to the momentum $p$ of the condensate.  The
sound velocity in the condensate at rest $(p=0)$ is given by
$c=\sqrt{gn/m}$.  When the condensate momentum $p < p_B\equiv mc$,
$\omega_q\neq 0$ for all $q$, while for $p > p_B$ there exists a value
of the momentum $q$ for which $\omega_q = 0$.  The time evolution of
the coefficients $c_q(t)$ can be calculated analytically following the
quasi-particle expansion approach developed in \cite{morgan}.  We
obtain

\begin{equation}
|c_q(t)|^2 = 
\left\{
\begin{array}{l c}
 \sqrt{ \frac{q^2}{q^2+ 4mgn} } |I_q|^2
\frac{\sin^2(\omega_q t)}{\hbar^2\omega_{q}^2} ; & \omega_q\neq 0 \\ 
 \sqrt{ \frac{q^2}{q^2+ 4mgn} } |I_q|^2 
\frac{t^2}{\hbar^2} ; & \omega_q = 0
\end{array}
\right.
\label{qpa}
\end{equation}
being $I_q=\frac{1}{L}\int dx e^{-iqx/\hbar}V_{def}(x)$ the spatial Fourier 
transform of the defect. 
According to Eq.(\ref{qpa}), when $\omega_q\neq 0$, the
quasi-particles modes are {\it stable}: their amplitudes oscillate in
time remaining finite. Instead, the amplitude of the {\it unstable}
mode with $\omega_q=0$ explodes, leading to a strong dissipation of
the condensate kinetic energy.

It is worth emphasizing here the main physical differences between
energetic and dynamical instabilities.  The latter is characterized by
a complex dispersion relation so that, even in absence of perturbing
potentials, small fluctuations in the population of the quasi-particle
modes increase exponentially fast. On the contrary, when the system is
energetically unstable, fluctuations remain small unless the
condensate is perturbed by an external defect.  In this case,
Eq.(\ref{qpa}) predicts an increase of the instable mode quadratically
in time.

We notice that in the Bogoliubov linear approach, the Landau critical
velocity is indipendent of the shape and the strength of the defect.
With a $\delta$-like defect $V_{def}=V_0\delta(x/L)$, $I_q = V_0$ for
all $q$, so that all quasi-particle modes are excited.  Instead, with
a periodic potential $V_{def}=V_0 \cos(\alpha x/\hbar)$, $I_q= (V_0/2)
\delta_{q=\pm \alpha}$ and it is possible to selectively excite the
two modes at $q=\pm \alpha$ \cite{note}.  In the Bogoliubov
framework it is possible to predict some simple typical behaviours:

1) for $p\neq 0$ the spectrum becomes asymmetric and the dynamics,
involving the two modes at $q=\pm \alpha$, will be dominated by the
the quasi-particle mode with smaller frequency, since its amplitude is
enhanced by $\omega_q$ at the denominator;

2) this quasi-particle excitation contains two momentum components at
$\pm q$, the relative weight of which depends on $\alpha$: for
small $\alpha$ the phonon character of the excitation will be strong
($U_q \sim V_q$) and both momentum components become important; for
large $\alpha$ the excitation is almost free-particle like and it
mostly contains only one momentum component at $q$.

From the above considerations, one can deduce that for large $p$ and
large $\alpha$ only a single momentum component other than the
condensate will be considerably excited, permitting to consider a
two-mode approximation, as we will discuss in the following.

To inquiry about the energetical stability of the system using a
spatially periodic defect, it is necessary to scan the wavevector of
the periodic potential $\alpha$ over all possible values. The Landau
criterion can be then stated in the following terms: {\it below the
critical velocity, the depletion from the condensate oscillates with
finite amplitude for all $\alpha$; above the critical velocity, there
exist a specific value of $\alpha$ at which the depletion grows
quadratically in time}.  It is interesting to point out that a
$\delta$-like defect exciting simultaneaously all momenta plays the
role of the scan over all possible $\alpha$ mentioned just
above. However, this is of course true only in the linear regime.  In
the non linear regime mode-coupling makes things much more complicated
and of difficult interpretation, so that we restrict ourselves to
consider in detail only the case of a spatially periodic defect.

{\it Numerical Simulations.}  
Bogoliubov's predictions can be verified numerically by integrating
the full GPE.  The numerical solutions of the GPE are well 
approximated by Bogoliubov theory until the amplitude of the depletion
is of the order of few percents of the condensate. 
At longer times, the two evolutions differ dramatically.  
First of all, as expected, also the population of the unstable mode 
oscillates, since it cannot grow without limit.
However stable modes might oscillate with larger amplitude than the
unstable one, and in general they oscillate with periods and
amplitudes different from those predicted by Bogoliubov and strongly
depending on the value of $V_0$.  The main consequence is that the
clear signature of the onset of energetic instability present in
Bogoliubov theory does not exists: for each value of the momentum of
the condensate $p$ and each value of the wavevector $\alpha$ of the
perturbing potential, the excited modes oscillate in time. It is not
possible to claim whether for a given $p$ and a given $\alpha$ the
system is stable or unstable, but a comparison of the
results at different $\alpha$ is needed.

\begin{figure}[h]
\begin{center}
        \includegraphics[width=0.8\linewidth]{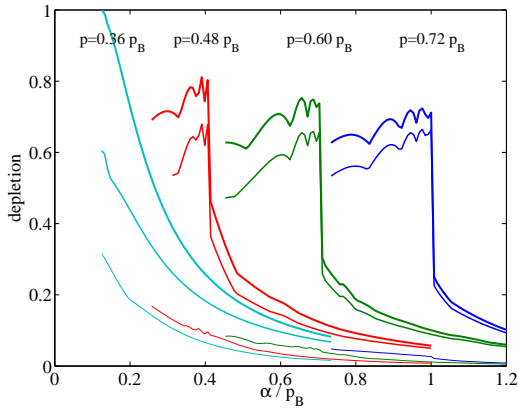}
\end{center}
\caption{Depletion and maximum amplitude of oscillation of the
  momentum component at $q=\pm \alpha$ for $p=0.36, 0.48, 0.6, 0.72\;
  p_B$ and $V_0=0.2 \;gn$. The thick line is the depletion from the
  condensate, while the normal and thin lines below are the
  contributions to it from the momentum component at $q=-\alpha$ and
  $q=+\alpha$ respectively.}
\label{fig1}
\end{figure}

In order to identify a general signature of the energetic instability,
we performed several numerical simulations fixing the momentum of the
condensate $p$ and changing the wavevector $\alpha$ and the strength
$V_0$ of the defect.  In Fig.~\ref{fig1}, we report the depletion from
the condensate and the population of the momentum components at
$q=\pm\alpha$ as a function of $\alpha$ for different values of
$p$. Comparing the results relative to different momenta $p$, we
observe different situations. For large $p$, the depletion presents a
discontinuity at a certain value of $\alpha$. Below a critical value
$p_c$ this discontinuity disappears. We define operationally the
critical velocity for energetic instability as the smallest velocity
of the condensate at which we observe such a discontinuity in the
depletion.  The critical value strongly depends on the strength of the
defect $V_0$, and, decresing $V_0$, the depletion asymptotically
reproduces the Bogoliubov prediction.

At this point we can remark the relation with the previous
formulation of the Landau criterion: {\it below the critical
velocity, the depletion varies smoothly over all $\alpha$; above the 
critical velocity, there exist a specific value of $\alpha$ at which
the depletion makes a jump},  which we take as the signature of 
the energetic instability.
In the limit $V_0 \to 0$, the region around the discontinuity becomes
sharper and asimptotically approaches the Bogoliubov prediction.

\begin{figure}[h]
\begin{center}
        \includegraphics[width=0.8\linewidth]{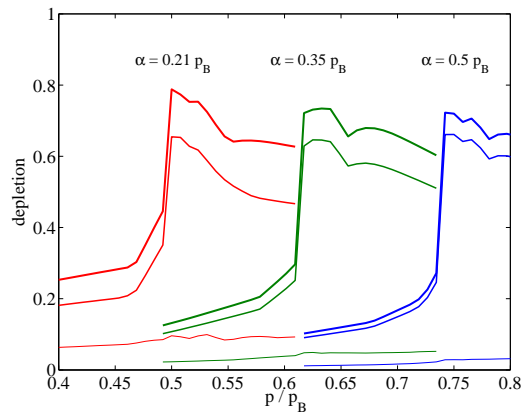}
\end{center}
\caption{Depletion and maximum amplitude of oscillation of the
  momentum component at $q=\pm \alpha$ as a function of $p$ for
  $\alpha=0.21, 0.35, 0.5 \;p_B$ and $V_0=0.2 \;gn$. The thick line is
  the depletion from the condensate, while the normal and thin lines
  below are the contributions to it from the momentum component at
  $q=-\alpha$ and $q=+\alpha$ respectively.}
\label{fig2}
\end{figure}

From the experimental point of view, it would be easier to study 
the response of the system at a fixed $\alpha$ as a function of 
the condensate momentum $p$ as shown in Fig.~\ref{fig2}.
The jump in the depletion occurs at a momentum larger than the critical 
one, and it is a signature of the energetic instability of the system 
at such value of $p$.

\begin{figure}[h!]
\begin{center}
\includegraphics[width=0.8\linewidth]{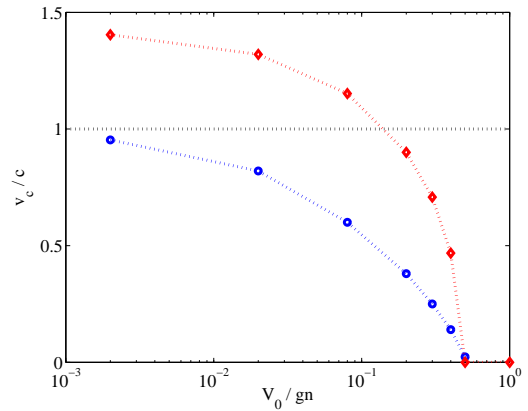}
\end{center}
\caption{Critical velocity as a function of the strength of the defect
 as defined in the text. The blue dots are the values obtained by GP
 simulations, the diamonds are obtained from the two-mode effective
 potential and the dotted line indicates the sound velocity $c$. }
\label{critical_vel}
\end{figure}

Following our definition, the critical velocity can be studied as a
function of $V_0$, as shown in Fig.~\ref{critical_vel}.  For small
$V_0$, we recover the Landau critical velocity, indicated by the
dotted line at $v=c$.  The limit of very small defect ($V_0<10^{-3}
gn$) can not be investigated numerically due to the very long time
scale of the dynamics.  The critical velocity decreases with
increasing $V_0$ and practically vanishes around $V_0 = gn/2$. For
$V_0>gn/2$, being the strength of the defect larger than the
interaction energy, the dynamics involves a huge number of
excitations, and it is difficult to establish numerically the presence or the
absence of the jump. 

{\it Nonlinear two-mode model:} As we have briefly mentioned above,
there is a regime (large $p$, large $\alpha$) in which a part from the
condensate only one single momentum component is considerably
populated and one can develop a two-mode model. 
We write the order parameter as

\begin{equation} 
\label{ansatz}
\psi(x,t)=  \frac{1}{\sqrt{L}}  \sum_{k} a_k(t)e^{ikx/\hbar},
\end{equation}
with $\sum_k|a_k(t)|^2=1$.  The initial conditions are $a_{k=p}=1$ and
$a_{k \neq p}=0$.  Replacing (\ref{ansatz}) in the GPE, we obtain a
set of coupled equations for the amplitudes $a_k(t)$, the coupling
between different modes being provided by the presence of the defect
and being proportional to its strength $V_0$.  The system of equations
shows that the defect couples the condensate to the modes with
momentum $k=p\pm \alpha$, and to higher harmonics, with momentum $k= p
\pm \ell \alpha$, with $\ell$ arbitrary integer.  A drastic
simplification in this set of coupled equations is obtained noticing
that, when both the momentum of the condensate $p$ and $\alpha$ are
large, only the momentum $k = p - \alpha$ (for $p>0$) couples
significantly to the condensate. 
This is consistent to the picture above, which predicted that, 
for $p$ and $\alpha$ large, the Bogoliubov mode with
$q=-\alpha$ lying in the free-particle region of the spectrum
($U_q\simeq 1$ and $V_q\simeq 0$) dominates the dynamics and gives
rise to a single momentum component at $p-\alpha$. 

The equation of motion for the relative population of the two modes
$z=(N_p-N_{p-\alpha})/N \in  [-1,1]$ is

\begin{equation} \label{equation_motion}
\dot{z}(t)^2 + W[z(t)]= const.,
\end{equation}
with

\begin{eqnarray}
W(z)= \frac{- V_0^2 (1-z^2) + [\Delta E (1 -z) - gn(1-z^2)/2]^2}{\hbar^2}
\end{eqnarray}
where $\Delta E=p^2/2m-(p-\alpha)^2/2m$, and the initial condition are
$z_0=z(0)=1$ and $\dot{z}(0)=0$.  These equations are similar to the
Josephson equations for the double well and many other two-mode
nonlinear systems ~\cite{joseph_double_well}.  With a mechanical
analogy, the dynamics of the system is described by the motion of a
particle in the potential $W(z)$ in Eq. (\ref{equation_motion}).  When
the condensate has momentum $p$ smaller than the critical value, the
effective potential displays a double-well shape for every value of
$\alpha$. The potential barrier remains higher than the initial energy
of the system so that the motion is always localized in the right well.
On the other hand if the momentum $p$ of the condensate is larger than
the critical value, by decreasing $\alpha$ the potential $W(z)$
displays a transition from a double-well to a single-well, i.e. the
potential barrier becomes smaller than the initial energy of the
system. In this case, the population of the excited mode suddenly
increases, qualitatively reproducing the jumps obtained with the GP
simulation shown in Figs.(\ref{fig1},\ref{fig2}).

\begin{figure}[h!]
\begin{center}
\includegraphics[width=0.8\linewidth]{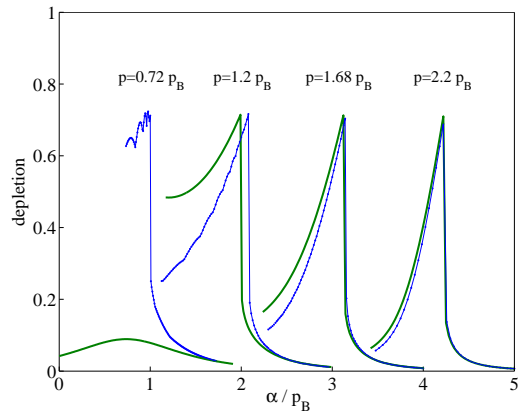}
\end{center}
\caption{Comparison between the results of the two-mode model (green
lines) with the results of the GP equation (blue lines with dots).}
\label{compare2mode}
\end{figure}

We have checked the capability of the two-mode model of reproducing
the results of the GP simulations, and confirmed that it works well for
large $p$ and large $\alpha$, as shown in Fig.~\ref{compare2mode}.

Close to the critical velocity, the two-mode model does not work
quantitatively well and overestimates the critical velocity, as shown
in Fig.~\ref{critical_vel}. However it reproduces the decrease of the
critical velocity as a function of $V_0$ and gives correctly the point
at which the critical velocity becomes negligible.  Based on this
result, we can use the two-mode model to calculate the critical defect
strength, at which the critical velocity vanishes.  To that aim, we
consider small $p$ and small $\alpha$, i.e. small $\Delta E$. Within
this assumption the effective potential can be expanded and becomes
 
\begin{eqnarray}
W(z) &\approx& -V_0^2(1-z^2)+(gn/2)^2(1-z^2)^2+\\
 &&-gn \Delta E (1-z^2)(1-z). \nonumber
\end{eqnarray}
When $V_0=gn/2$, the effective potential undergoes a double to 
single-well transition for each value of $p$: 
for $\alpha>2p$, $\Delta E <0$ and we have that $W(0)>W(1)$; 
for $\alpha\in [0,\; 2p]$, $\Delta E >0$ and we obtain that $W(0)<W(1)$.  
This means that at $\alpha\approx 2p$, $W(z)$ undergoes a transition 
from double to single-well,  and hence the population of the excited mode 
undergoes a jump, signature of the energetic instability.

{\it Conclusions.}  The Landau criterion for superfluidity predicts
that a weakly interacting Bose-Einstein condensate can flow without
dissipation when its group velocity is smaller than the sound
velocity.  We found that the Landau criterion remains valid only in
the linear (Bogoliubov) regime or, in other words, for vanishingly
small perturbations. Indeed, the superfluidity critical velocity
rapidly decreases while increasing the strength of the defect, up to a
critical value of the perturbation where the flow is always
dissipative. This finding can be tested experimentally with spatially
periodic perturbations, and can qualitatively explain the critical
velocities lower than sound observed with condensates flowing against
large defects.

{\it Aknowledgements.} We thank F. Dalfovo and W.D. Phillips for useful 
discussions. This work has been partially supported by the DOE.


\begin{thebibliography}{9} 

\bibitem{Huang} Kerson Huang 
  \emph{Statistical Mechanics}. John Wiley $\&$ Sons, Second edition.

\bibitem{pita} G. E. Astrakharchik and L. P. Pitaevskii,
Phys. Rev. {\bf A 70}, 013608 (2004) 

\bibitem{leggett} For a review see A.J. Leggett, Rev. Mod. Phys. {\bf 73}, 307 (2001) 

\bibitem{Abo-Shaeer} J.R. Abo-Shaeer et al., Science {\bf 292}, 476 (2001).

\bibitem{Hodby} E. Hodby et al., Phys. Rev. Lett. {\bf 88}, 010405 (2002).
  
\bibitem{Davis} K.B. Davis et al., Phys. Rev. Lett. {\bf 75}, 3969 (1995).
  
\bibitem{Marago_scissors} O. Marag\`o et al., Phys. Rev. Lett. {\bf 86}, (2001).
 
\bibitem{jose} Cataliotti et al., Science {\bf 293} 843 (2001); 
M. Albiez et al., Phys. Rev. Lett. {\bf 95}, 010402 (2005) 

\bibitem{onofrio} R. Onofrio et al., Phys. Rew. Lett.{\bf 85}, 2228 (2000).

\bibitem{Jackson} B.Jackson, J.F.McCann and C.S.Adams, Phys. Rev. 
{\bf A 61}, 013604 (2000), {\bf 61}, 051603(R) (2000).
  
\bibitem{tilley} D. R. Tilley and J. Tilley, Superfluidity and
Superconductivity (Adam Hilger Ltd., Bristol/Boston,1991).

\bibitem{kozuma} M. Kozuma et al., Phys. Rev. Lett. {\bf 82}, 871
(1999); G. Birkl et al., Phys. Rev. Lett. {\bf 75}, 2823 (1995).

\bibitem{Stenger} J.Stengel et al., Phys. Rev. Lett. {\bf 82}, 4569 (1999).
 
\bibitem{davidson} J.Steinhauer et al., Phys. Rev. Lett.  {\bf 88}, 120407 (2002).

\bibitem{dalfovo} C. Tozzo and F. Dalfovo, New J. Phys., {\bf 5}, 54 (2003).

\bibitem{morgan} S.A. Morgan, S. Choi, K. Burnett, and M. Edwards,
 Phys. Rev. {\bf A57}, 3818 (1998); P.B.Blakie, R.J.Ballagh and
 C.W. Gardiner, Phys. Rev. {\bf A65}, 033602 (2002).
 
\bibitem{note} With respect to the experiments performed with the
Bragg potential, where the condensate is at rest and the periodic
potential $V_B \cos(q_B x -\omega_B t)$ moves with velocity
$\omega_B/q_B$, this is simply a change of reference frame: in our
case the potential is at rest and the condensate moves with velocity
$v=p/m$. Therefore the significant quantities, $q_B$ and $\omega_B$ in
the Bragg experiments, $\alpha$ and $p$ in our system, are related one
to the other in the following way $\alpha=\hbar q_B$ and $p=
m\omega_B/q_B$.
 
\bibitem{trombettoni} Compare this approach with the case of a defect
on teh nonlinear discrete Schroedinger equation on a ring,
A. Trombettoni, A. Smerzi and A.R. Bishop, Phys. Rev. Lett. {\bf 88}
173902 (2002)

\bibitem{joseph_double_well} Raghavan et al., Phys. Rew. {\bf A59}, 59
(1999) and ref.s therein.

\end{thebibliography}
\end{document}